# The SNO+ Experiment

M.C. Chen  (for the SNO+ collaboration)
*Department of Physics, Queen's University, Kingston, ON K7L 3N6, Canada*

The SNO+ experiment is the follow-up to the Sudbury Neutrino Observatory (SNO).  The heavy water that was in SNO will be replaced with a liquid scintillator of linear alkylbenzene (plus fluor).  SNO+ has many physics goals including detecting *pep* and *CNO* solar neutrinos, detecting geo-neutrinos, studying reactor neutrino oscillations, serving as a supernova neutrino detector and carrying out a search for neutrinoless double beta decay by adding neodymium to the liquid scintillator.  Since a large amount of $^{150}$Nd isotope can be added to SNO+, a competitive search would be possible, with sensitivity below 100 meV using natural Nd and sensitivity below 40 meV with enriched neodymium.

## 1.  INTRODUCTION

The Sudbury Neutrino Observatory (SNO) has concluded. The heavy water that was in the SNO detector has been removed and we plan to fill SNO with a liquid scintillator – this project is SNO+.  With a liquid scintillator, the light yield in the detector increases by a factor of ~50, allowing SNO+ to study neutrino physics at lower energies.  Solar neutrinos at lower energies are interesting as precision probes of neutrino physics and as a means of learning about solar physics, revisiting the intent of Ray Davis when he set out to detect solar neutrinos forty years ago.  The antineutrinos emitted by natural radioactivity in the Earth (uranium and thorium) can be detected by a large liquid scintillator detector.  By doing so, one can assay the Earth by looking at its neutrino emission, thereby providing constraints on the radiogenic portion of Earth's heat flow and on the radiochemical composition of the Earth's mantle and crust.  Nuclear power reactors in Ontario are farther from the SNO+ detector than the typical distance of reactors in Japan to the KamLAND experiment.  The signal from reactor antineutrinos would still be easily detected by SNO+ and spectral features observed in KamLAND due to neutrino oscillations would shift to higher energy in SNO+ (L/E is a constant) – the observation of which would demonstrate the oscillation phenomenon and allow for added constraints on oscillation parameters.  A large liquid scintillator detector serves as an excellent supernova neutrino monitor.

An additional aim of the SNO+ experiment is the search for neutrinoless double beta decay.  In SNO+ we also plan to deploy Nd-loaded liquid scintillator.  0.1% Nd loading is easily achieved and in a kiloton-sized detector this corresponds to adding of order 1 ton of Nd.  Even with natural Nd, the isotopic abundance of $^{150}$Nd is 5.6% and 1000 kg Nd thus has 56 kg of $^{150}$Nd.  NEMO-3, in comparison, has 37 g of $^{150}$Nd; a large Nd-loaded liquid scintillator provides a way to mount an experiment with a significantly larger amount of neodymium than has been possible in the past.

In order to convert SNO to SNO+, the following actions are needed.  In SNO, the acrylic vessel contained heavy water (surrounded by normal, light water) and thus needed to be held up.  SNO+ will have scintillator with density less than one inside the acrylic vessel which is surrounded by water.  The acrylic vessel in SNO+ will be buoyant.  The design and installation of a hold-down system of ropes for the SNO+ acrylic vessel is required.  For SNO+, the scintillator and its components must be procured.  In order to achieve the ultra-low backgrounds required for the physics goals of the experiment, a scintillator purification system must be built (in place of the heavy water purification that was used in SNO).  These are the three large items that require significant cost and/or development.  In addition, a few minor upgrades will be made: to the cover gas system in order to lower radon backgrounds; to the electronics and DAQ to improve the ability to handle the higher data rates that will be seen in SNO+ (due to the greater light yield and higher event rates at lower energies); and new calibration sources and systems will be built, appropriate for the new physics and new detector in SNO+. All told, the cost and scale of the work to realize the SNO+ experiment is relatively small. This is, of course, the *raison d'être* for SNO+. We are able to make new measurements and explore new physics on a short timescale, by reusing the SNO detector.

In the following sections, a few more details about the double beta decay sensitivity will be discussed. The solar neutrino physics will also be examined. Concluding remarks about the status of the experiment will be made.

## 2. SNO+ DOUBLE BETA DECAY

The motivation for double beta decay searches is the determination of the charge conjugation nature of the neutrino. It is not known whether neutrinos are Majorana or Dirac particles. If neutrinos are Majorana particles, then the exchange of light (finite mass) Majorana neutrinos provides a mechanism for neutrinoless double beta decay to occur with the rate of this process proportional to the effective Majorana neutrino mass squared

$$<m_\nu> = \sum_i m_i U_{ei}^2$$

where $m_i$ are the masses of the three light neutrino mass eigenstates and $U_{ei}$ are the mixing amplitudes between these mass eigenstates and the electron neutrino flavor eigenstate. Neutrino oscillation experiments have determined mass squared differences and these suggest certain ranges for the effective Majorana neutrino mass. In particular, if the inverted neutrino mass hierarchy is correct, atmospheric neutrino oscillation results suggest an effective Majorana neutrino mass just below ~50 meV [1]. This has become a target for next-generation double beta decay experiments and is within the range of sensitivity of several proposed experiments. Another goal of upcoming double beta decay experiments is certainly to examine the claim of observation of neutrinoless double beta decay in $^{76}$Ge [2].

An interesting isotope for the search for neutrinoless double beta decay, 0νββ, is $^{150}$Nd. The endpoint $Q_{\beta\beta}$ of $^{150}$Nd is 3.37 MeV (the second highest after $^{48}$Ca). Neodymium has the largest phase space factor of all the double beta decaying isotopes (e.g. it's a factor of 33 greater than the phase space factor for $^{76}$Ge). This implies that for the same Majorana effective neutrino mass, the 0νββ rate in $^{150}$Nd is faster. All of these advantages contribute in the desired manner to help a liquid scintillator double beta decay experiment. A liquid scintillator will have poorer energy resolution than an experiment with a Ge detector, for example, or a cryogenic bolometer. Thus a high $Q_{\beta\beta}$ is valuable in order to place the signal above backgrounds from natural radioactivity. At 3.37 MeV, the endpoint is above backgrounds from the highest energy gamma line at 2.6 MeV, and above backgrounds from radon which do not exceed 3.2 MeV. In a large liquid scintillator detector very low radioactivity backgrounds can be achieved. Neodymium is thus a good choice for the SNO+ double beta decay experiment.

A double beta decay experiment searches for 0νββ events at the endpoint. Usually this is done by looking for a peak from those events and excellent energy resolution is desired. If one employs an isotope with an endpoint above the energy of background gamma lines from natural radioactivity, energy resolution is not as crucial for distinguishing the 0νββ signal from other gamma lines because there aren't any. Thus, it becomes possible to fit for the spectral distortion introduced by 0νββ events at the endpoint in excess to the spectrum from 2νββ events that occur. Figure 1 illustrates this for the SNO+ detector loaded with 500 kg of $^{150}$Nd isotope, at an achievable energy resolution for a liquid scintillator detector, and for an effective Majorana neutrino mass of 150 meV (at the level of the claim in [2]).

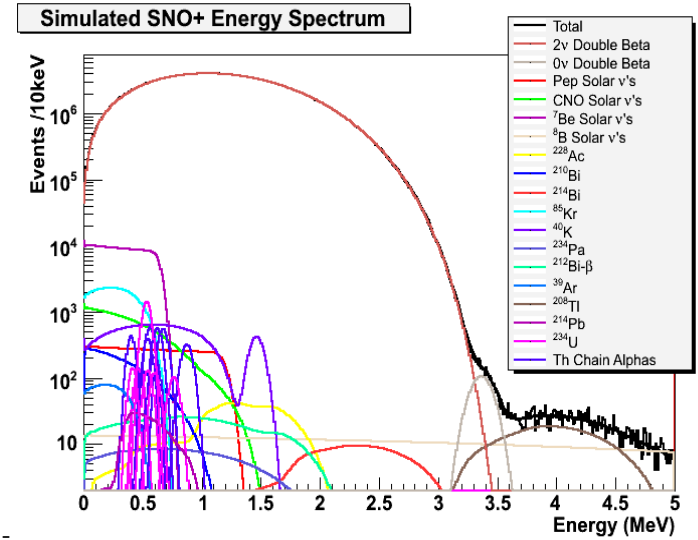

Figure 1: SNO+ energy spectrum (simulated one year) with 500 kg of $^{150}$Nd, energy resolution corresponding to light yield of 400 photoelectrons/MeV and an effective Majorana neutrino mass of 150 meV. The endpoint distortion is visible with high statistical significance. Only internal Th contamination and $^8$B solar neutrinos are backgrounds.

By fitting the endpoint spectrum shape one can extract the $0\nu\beta\beta$ signal or place limits on its contribution. In the spectrum simulated in Figure 1 backgrounds were simulated at the Th contamination levels achieved in Borexino [3]. The fast delayed coincidence (300 ns) between $^{212}$Bi-$^{212}$Po can be used to constrain the Th background at high energies. There exists the 3 minute alpha-alpha coincidence of $^{212}$Po-$^{208}$Tl that can directly tag these backgrounds to eliminate them, though the longer coincidence time makes using this more difficult. What remains irreducible though is the background from $^8$B solar neutrinos, though the $^8$B spectral shape can be inferred from fitting higher energy events.

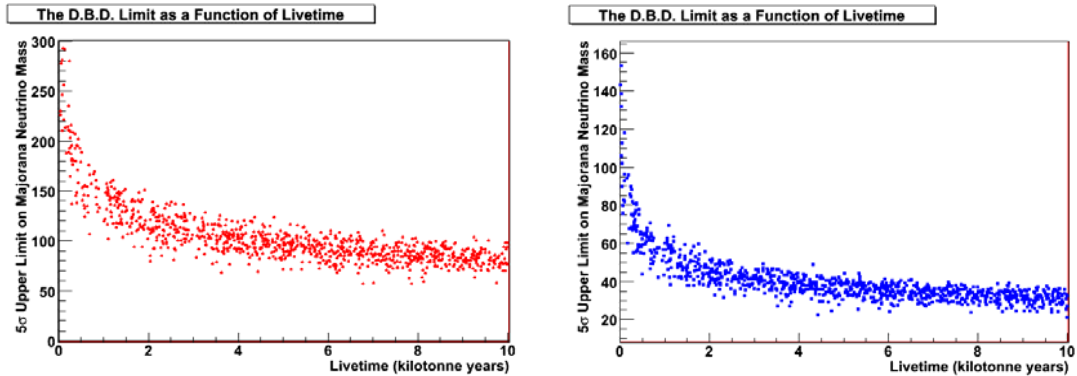

Figure 2: Sensitivity to neutrino mass [in meV] with natural Nd (left) and enriched Nd (right) in SNO+. These plots are frequentist-style determinations of the 5σ upper limit one would derive, where each point is an experiment.

Sensitivity studies have been made assuming different concentrations of isotope and are shown in Figure 2. For concentrations achievable with natural Nd, sensitivities (5σ) below 100 meV can be reached. For concentrations that would require isotopically enriched neodymium, sensitivities (5σ) reach down below 40 meV.

## 3. LOW ENERGY SOLAR NEUTRINOS

SNO+ aims to detect *pep* and *CNO* solar neutrinos. The flux of *pep* solar neutrinos is fundamental and related to the flux of *pp* solar neutrinos; it is calculated to better than ±1.5% uncertainty in the Standard Solar Model [4]. The cross section for neutrino-electron scattering has negligible uncertainty. A measurement of the rate of *pep* solar neutrino interactions can thus be used to determine the survival probability (for $E_\nu = 1.44$ MeV) precisely. Detecting the *pep* solar neutrinos and measuring their survival probability, with precision, is interesting from the point of view of new neutrino physics. The presence of phenomena such as non-standard interactions [5] or mass-varying neutrinos [6] might manifest itself as a different survival probability at the transition between vacuum dominance and matter dominance, in the oscillation of solar neutrinos. The transition region is at neutrino energies between 1-2 MeV. Measuring the survival probability of *pep* solar neutrinos has good sensitivity to new physics.

Detecting the *CNO* neutrinos would be interesting as it could reveal information relevant to solar elemental abundances. A recent problem in solar models is the fact that helioseismology seems incompatible with low metallicity solar models. Solar model sound speeds [4] no longer agree with helioseismology when elemental abundances from [7] are used. The *CNO* neutrino fluxes depend on these abundances; differences of 50-60% in the flux result from using the older metallicity values and the newer ones. A measurement of the *CNO* fluxes would suggest which is compatible.

## 4. SNO+ EXPERIMENTAL STATUS

All aspects of the experiment have been advanced, including: development of purification techniques for Nd and Nd-loaded scintillator, buckling and stress finite element analysis of the acrylic vessel hold-down rope net, engineering of the scintillator purification system, and development of a liquid scintillator based upon linear alkylbenzene (including stable Nd loading). The SNO+ experiment is ready to be implemented.

SNO+ has already received partial, but substantial, funding for final design/engineering and initial construction. A full capital proposal was submitted in October 2008. If approved, SNO+ will undertake construction activities in 2009 starting with the installation of the rope net that would hold down the acrylic vessel. Construction and installation of the scintillator purification and process systems would occur in 2010. We aim for the end of 2010 for the start of scintillator filling, and soon afterward the double beta decay phase of SNO+ with Nd.

## 5. PAPER SUBMISSION

### 5.1. Submitting to the arXiv Server

Authors using MS Word cannot submit source files to the ePrint arXiv server, but rather may need to submit a PDF file instead.

Submission to the arXiv server provides automated version control. If authors need to make changes, they can resubmit to the arXiv server, and a new version number is automatically applied, thus eliminating most version control problems. The ICHEP08 proceedings will automatically reference the latest version of the paper.

After the arXiv has accepted an author's paper, the author is requested to verify that the PDF the arXiv has is the same as their local PDF and that the format agrees with our example.

### 5.2. Submitting to the Conference

After the arXiv has accepted an author's paper and the author has verified that the arXiv's PDF is correct, the arXiv number – NNNN.NNNN (eg, 0809.0001) – needs to be entered on the ICHEP08 web site (http://www.hep.upenn.edu/ichep08/talks/proceedings/ ) so that it can be included in the official proceedings.